\documentclass[preprints,article,accept,moreauthors,pdftex,10pt,a4paper]{Definitions/mdpi} 
\firstpage{1} 
\pubvolume{xx}
\issuenum{1}
\articlenumber{5}
\pubyear{2018}
\copyrightyear{2018}
\history{Received: date; Accepted: date; Published: date}

\Title{Development of a neutron imaging station at the n\_TOF facility of CERN and applications to beam intercepting devices}

\Author{Federica Mingrone $^{1,2}$*, Marco Calviani $^{1,}$* \orcidA{}, Claudio Torregrosa Martin $^{1}$ \orcidB{}, Oliver Aberle$^{1}$, Michael Bacak$^{1,3,4}$, Enrico Chiaveri$^{1,5,6}$, Elvis Fornasiere$^{1}$, Antonio Perillo-Marcone$^{1}$, Vasilis Vlachoudis$^{1}$ and the n\_TOF Collaboration}

\AuthorNames{Federica Mingrone, Marco Calviani, Claudio Torregrosa Martin, Oliver Aberle, Michael Bacak, Enrico Chiaveri, Elvis Fornasiere, Antonio Perillo-Marcone and Vasilis Vlachoudis}

\address{%
$^{1}$ \quad European Organization for Nuclear Research, CERN, Geneva CH-1211, Switzerland; \\
$^{2}$ \quad National Technical University of Athens, Greece \\
$^{3}$ \quad Technische Universit\"at Wien, Austria \\
$^{4}$ \quad CEA Saclay, Irfu, Universit\'e\ Paris-Saclay, Gif-sur-Yvette, France \\
$^{5}$ \quad University of Manchester, United Kingdom \\
$^{6}$ \quad Universidad de Sevilla, Spain}

\corres{Correspondence: federica.mingrone@cern.ch, marco.calviani@cern.ch}

\abstract{A neutron radiography testing station has been developed exploiting the neutron beam of CERN's n\_TOF Experimental Area 2, located at the shortest distance to the neutron producing-target. The characteristics of the n\_TOF neutron beam for the imaging setup are presented in this paper, together with the obtained experimental results. The possible developments of neutron imaging capabilities of the n\_TOF facility in terms of detection-systems and beam-line upgrades are as well outlined.}


\begin{document}


\section{Introduction}
\label{sec:intro}

Neutron imaging is a well-developed non-destructive analysis method, exploited for a variety of applications from material science to aerospace technologies~\cite{BRENIZER201310,WORACEK2018141,jimaging3040052}. Due to their peculiar interaction with matter, neutrons act as probes penetrating thick-walled samples and providing an image of the transmitted radiation, which intensity depends on the thickness of the material layers and on the specific attenuation properties of that material. In this regards, neutron radiography can be assimilated to X-rays. However, while X-rays are attenuated more effectively by heavier materials like metals, neutrons allow to image light materials, such as hydrogenous substances, with high contrast, making the two imaging methods complementary to investigate the properties of object internal structures.

Several dedicated facilities are operating worldwide in order to provide high performance neutron imaging stations, in particular at research nuclear reactors and at spallation sources~\cite{jimaging3040052}. 

A neutron radiography testing station has been recently developed~\cite{Calviani:1979249,Mingrone:2241279} exploiting the neutron beam of the n\_TOF Experimental Area 2, located at the shortest distance from the spallation target~\cite{Sabate-Gilarte2017}, successfully inspecting the inner structure of several targets previously irradiated at the HiRadMat facility of CERN~\cite{HiRadMatFac} and of a spent antiproton production target used in the Antiproton Decelerator
(AD) target area of CERN. The characteristics of the n\_TOF neutron beam for the imaging setup will be here presented, together with the results obtained. The possible further developments of neutron imaging capabilities of the n\_TOF facility in terms of detection-systems and beam-line upgrades will be as well outlined.
 
\section{The n\_TOF facility}
The neutron Time-Of-Flight facility of CERN, called n\_TOF, started its operation in 2001, based on an idea by C. Rubbia \cite{Rubbia}, providing a very high-resolution neutron energy spectrometer with the aim of measuring neutron-induced reaction cross sections. The facility ran from 2001 to 2004 (n\_TOF Phase-1) and, after a four years stop it resumed operation at the end of 2008 till the end of 2012 (n\_TOF Phase-2). During the Long Shutdown 1 (LS1) of CERN a second short 20 m flight-path~\cite{EAR2}, complementing the existing 185~m one, has been constructed from May 2013 and completed in July 2014, starting the n\_TOF Phase-3.

In the facility, neutrons are produced by spallation reactions induced by a 20~GeV/c pulsed proton-beam from the CERN Proton Synchrotron on a water-cooled lead target. The pulsed neutron source is operating together with a moderation system, consisting of a 4 cm layer of borated and demineralized light water for the first and the second experimental areas, respectively, with the purpose of widening the neutron energy spectrum down to the epithermal and thermal regions and to enhance the neutron-energy resolution. In this way, the n\_TOF neutron beam covers about eleven orders of magnitude in energy from thermal to GeV in the first experimental area (EAR1), and to hundreds of MeV in the second experimental area (EAR2). The aim of borated water is to suppress the in-beam 2.2 MeV $\gamma$-ray background coming from thermal neutron capture in hydrogen.

\begin{figure}[b]
\centering 
\includegraphics[width=.6\textwidth]{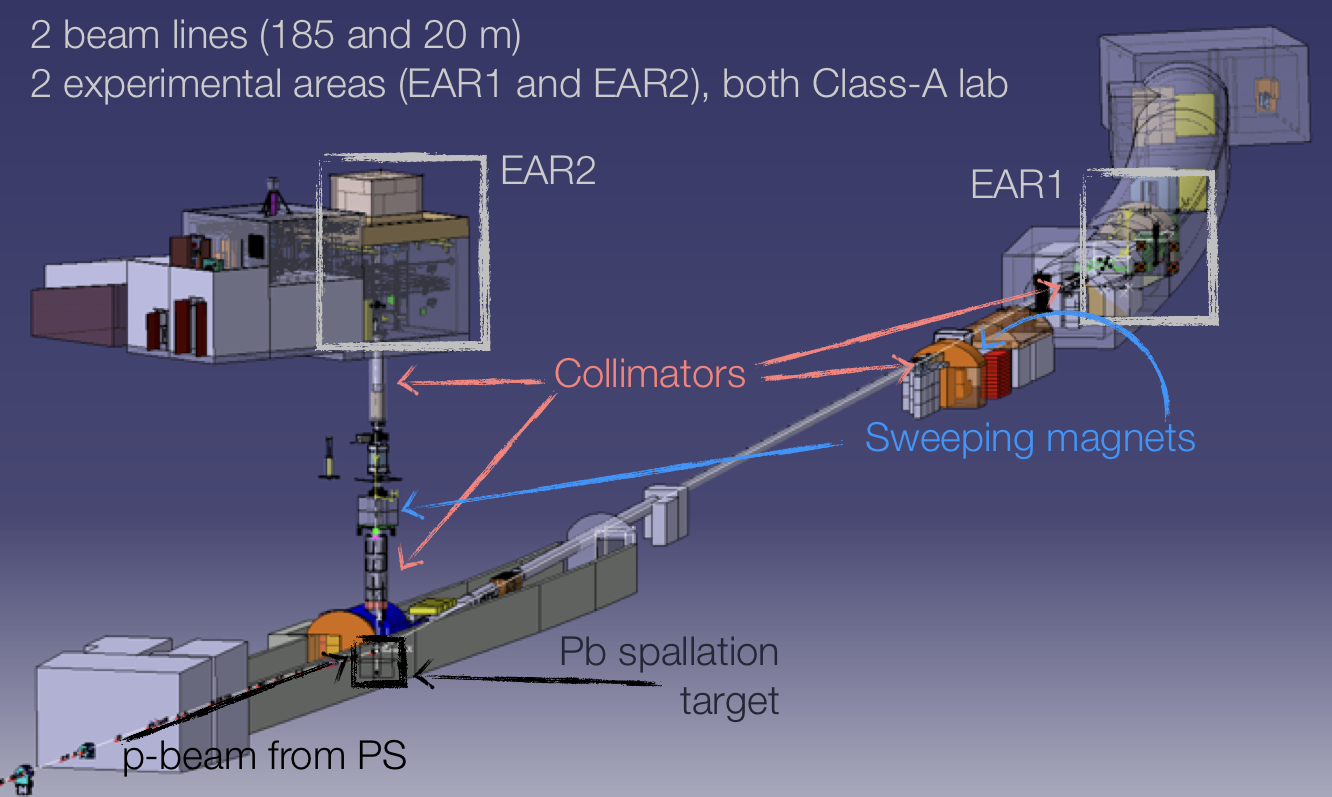}
\caption{\label{fig:beamline} Scheme of the n\_TOF facility. The main component of the neutron source and the two neutron beam-lines are highlighted.}
\end{figure}

From the target two beam lines start: the first, 185-m long and leading to the experimental area 1 (EAR1), is going along in the horizontal plane with an angle of 10$^\circ$ with respect to the impinging proton beam, while the other extends 20 m vertically on top of the target before reaching the second experimental area (EAR2). Both lines are equipped with two collimators each, of which the second is located immediately before the entrance of the experimental areas to give the neutron beam its final shape. Located in between the two collimators a so-called sweeping magnet is deviating the remaining charged particles that are travelling along the neutron beam. In Figure~\ref{fig:beamline} a scheme of the beam-lines and their elements is shown.

\begin{table}[ht]
\centering
\caption{\label{tab:nTOF} Features of the n\_TOF neutron beam in the two available experimental areas. Data extracted from Refs.~\cite{guerrero,EAR2,Sabate-Gilarte2017}.}
\smallskip
\begin{tabular}{|l|ll|}
\hline
      & EAR1  & EAR2    \\
\hline
Neutron flux (n/cm$^2$/pulse)   & about $2 \times 10^5$  & about $5 \times 10^6$ \\
Energy range & thermal to 1 GeV & thermal to a few hundreds MeV    \\
Energy resolution & $\Delta E/E = 10^{-4}$ at 1 eV & $\Delta E/E = 10^{-3}$ at 1 eV \\
\hline
\end{tabular}
\end{table}

\begin{figure}[hb]
\centering
\includegraphics[width=.6\textwidth]{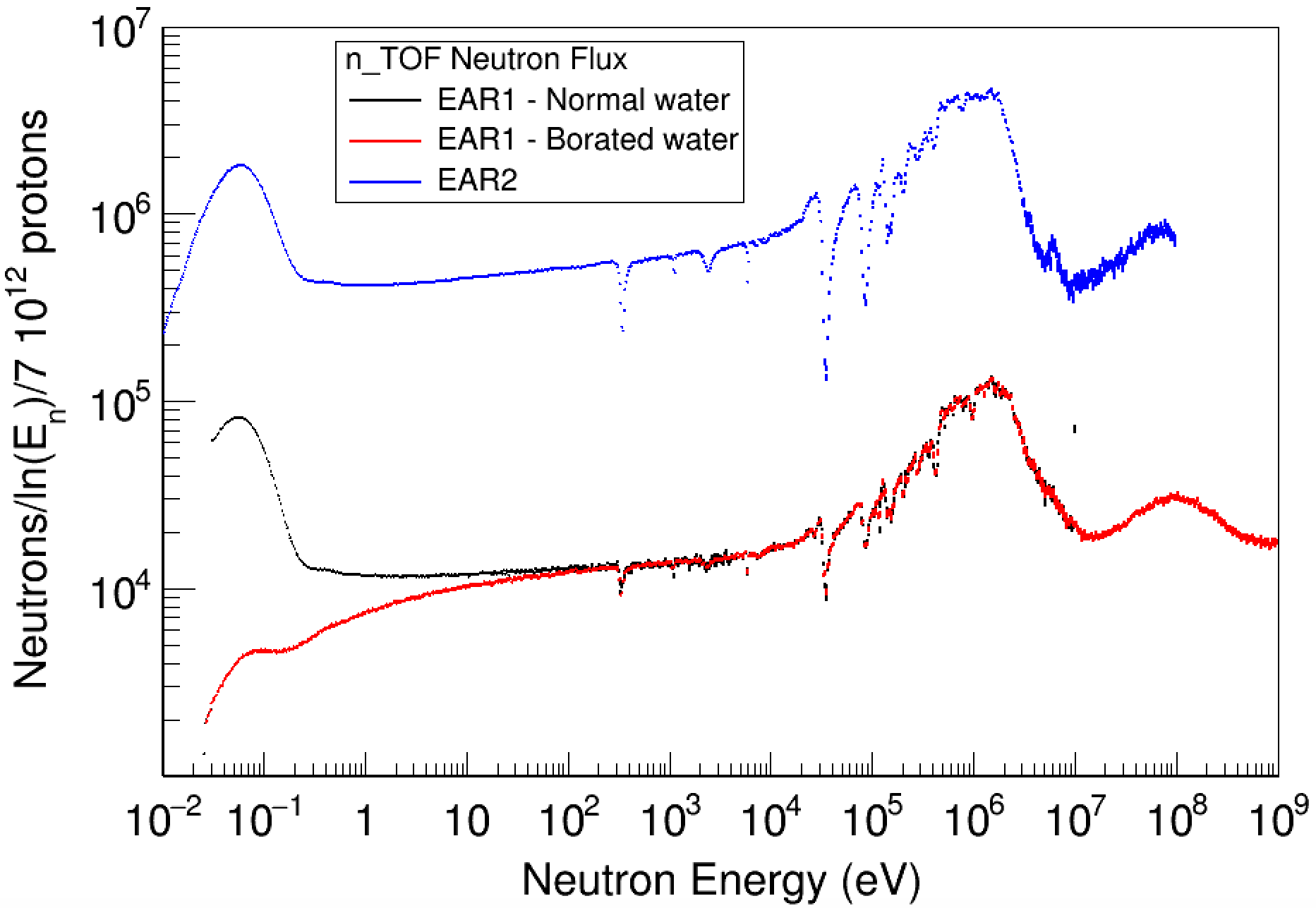}
\caption{\label{fig:flux} Neutron flux in EAR1 (black and red lines) and EAR2 (blue line) as a function of the neutron energy. In the low neutron energy region the difference resulting in using normal (black) or borated (red) water can be seen.}
\end{figure}

In Table~\ref{tab:nTOF} the main characteristics of the neutron beam as arriving in the two experimental areas are listed. As can be inferred, the two areas have complementary features. If in fact EAR1 is best suited for high-resolution measurements, particularly in the resolved resonance region, and to extend cross-sections to the high neutron energies ($E_n>100$ MeV), the exceptionally high instantaneous intensity of the neutron beam in EAR2, with about $5\times10^6$ neutrons per 30 ms-long bunch, allows to measure cross-sections of very low mass samples ($< 1$ mg), radioisotopes with short half life, and reactions with very small cross-sections. Recently, the second experimental area has been exploited as a neutron imaging facility, counting on a fluence of about 10$^6$ neutrons in the thermal energy range (E$_n<1$~eV). A comparison of the neutron fluence in the two experimental areas is presented in Figure~\ref{fig:flux}, where also the difference due to the choice of borated or demineralized water as moderation liquid for EAR1 can be seen.

\section{Experimental setup}
Conventional neutron-imaging techniques are based on mapping the attenuation of a neutron beam when transmitted through a sample. The resulting intensity map can be represented as an image with two main parameters, spatial resolution and contrast. To assess the quality of a neutron radiography station different parameters should be considered~\cite{imaging}:
\begin{itemize}
    \item Neutron fluence at the exit of the collimator pinhole, which determines the signal to noise ratio of the neutron radiography.
    \item Neutron fluence at the sample position, which, for a simple beam-line configuration consisting of just one collimator, depends on the diaphragm or collimator diameter D, the intensity at the diaphragm $\Phi_S$ and the collimator-sample distance L via the relation $\Phi_0 = \Phi_S / (4L/D)^2$. However, as described in Section \ref{ssec:beam_par}, the beam-line of the n\_TOF facility presents a more complex design, and therefore Monte Carlo simulations are required to correctly estimate the neutron flux at the sample position. 
    \item Area at the sample position with an homogeneous neutron flux, which can be evaluated from D and L via geometrical optics.
    \item Geometric blur, which can be expressed by $d=l/(L/D)$, where $l<L$ is the distance between the sample  position and an imaging device.
    \item Magnification factor, $M=L/(l-L)$, particularly critical for tomography as it requires to treat different l depending on the point of interest.
\end{itemize}

\subsection{n\_TOF Experimental Area 2 (EAR2) neutron beam parameters} \label{ssec:beam_par}

\begin{figure}[ht]
\centering 
\includegraphics[width=.52\textwidth]{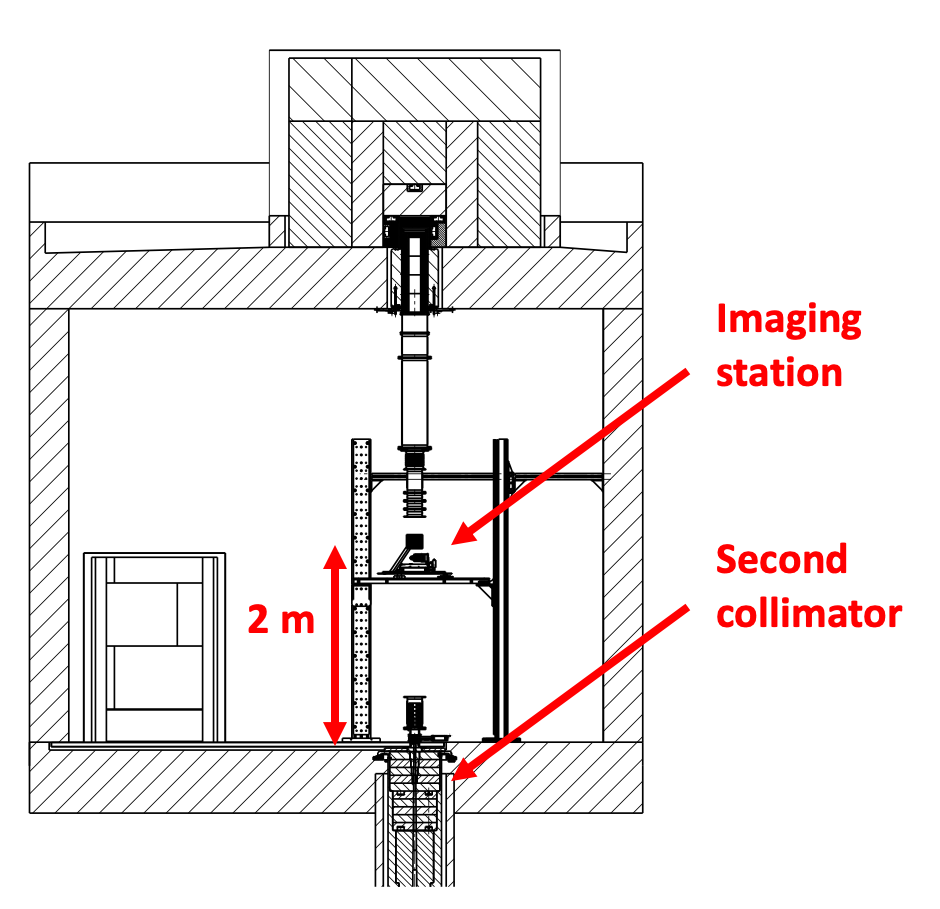}
\qquad
\includegraphics[width=.4\textwidth]{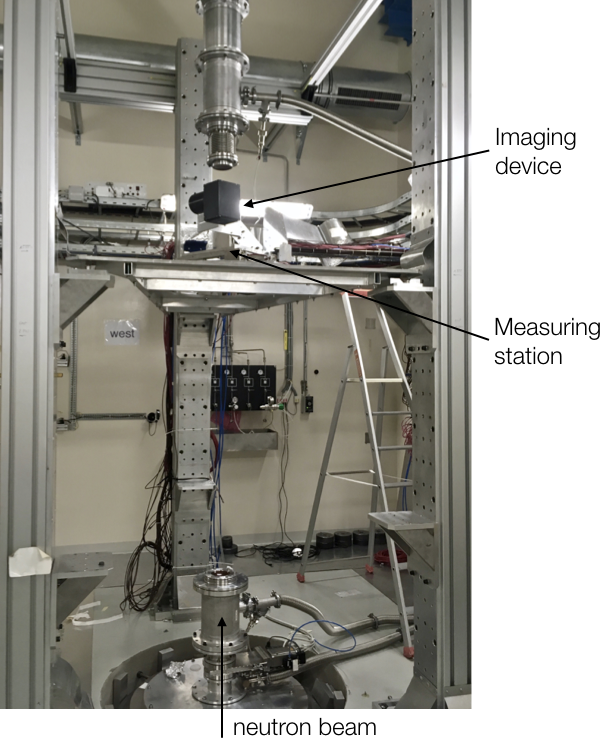}
\caption{\label{fig:imaging_station} Scheme (left) and picture (right) of the n\_TOF Experimental Area 2 bunker setup for neutron imaging measurements.}
\end{figure} 

The n\_TOF EAR2 vertical beam-line is designed to have a 1-m long first collimator with 20-cm diameter, positioned at a distance of 7.5 m with respect to the spallation target. The second collimator is located 20 cm below the floor of EAR2. There are two versions of the collimator, both are 3 m long and have a conical inner shape, with a larger aperture of 70 mm for the small and 96.8 mm for the big one, and a smaller determining the beam profile of 21.8 mm and 60 mm, respectively. Without any modification of the beam line the diameter of the collimator inlet aperture $D$ can, therefore, be chosen as 70 mm and 96.8 mm. This parameter given, it is possible to adjust both $l$ and $L$ in the experimental area to optimize the achievable resolution: the distance $l$ between the sample and the detector was minimized to avoid damaging the scintillator layer in the operations, ranging from 50 to 100 mm. The sample was put 2 m above the floor of experimental area, as the best compromise between the best resolution and the accessibility of the imaging station, resulting in a total sample-collimator distance of about 5 m. The setup of the EAR2 bunker for imaging measurements is shown in Figure~\ref{fig:imaging_station}.

The characteristics of the neutron beam have been evaluated through FLUKA Monte Carlo code~\cite{bib:Fluka1,bib:Fluka2} simulations. At the experimental location - 220 cm above the end of the collimator - and in the optimistic case of 1 pulse every 1.2 seconds, about $8\times10^5$ and $5\times10^5$ thermal neutrons/cm$^2$/pulse obtained at the sample with the big and small collimator, respectively. The flat part of the beam profile is expected between 4 and 6 cm diameter for the small collimator, and between 9 and 11 cm diameter for the big one, the difference depending on the sensitivity to the beam halo.

\begin{figure}[ht]
\centering 
\includegraphics[width=.48\textwidth]{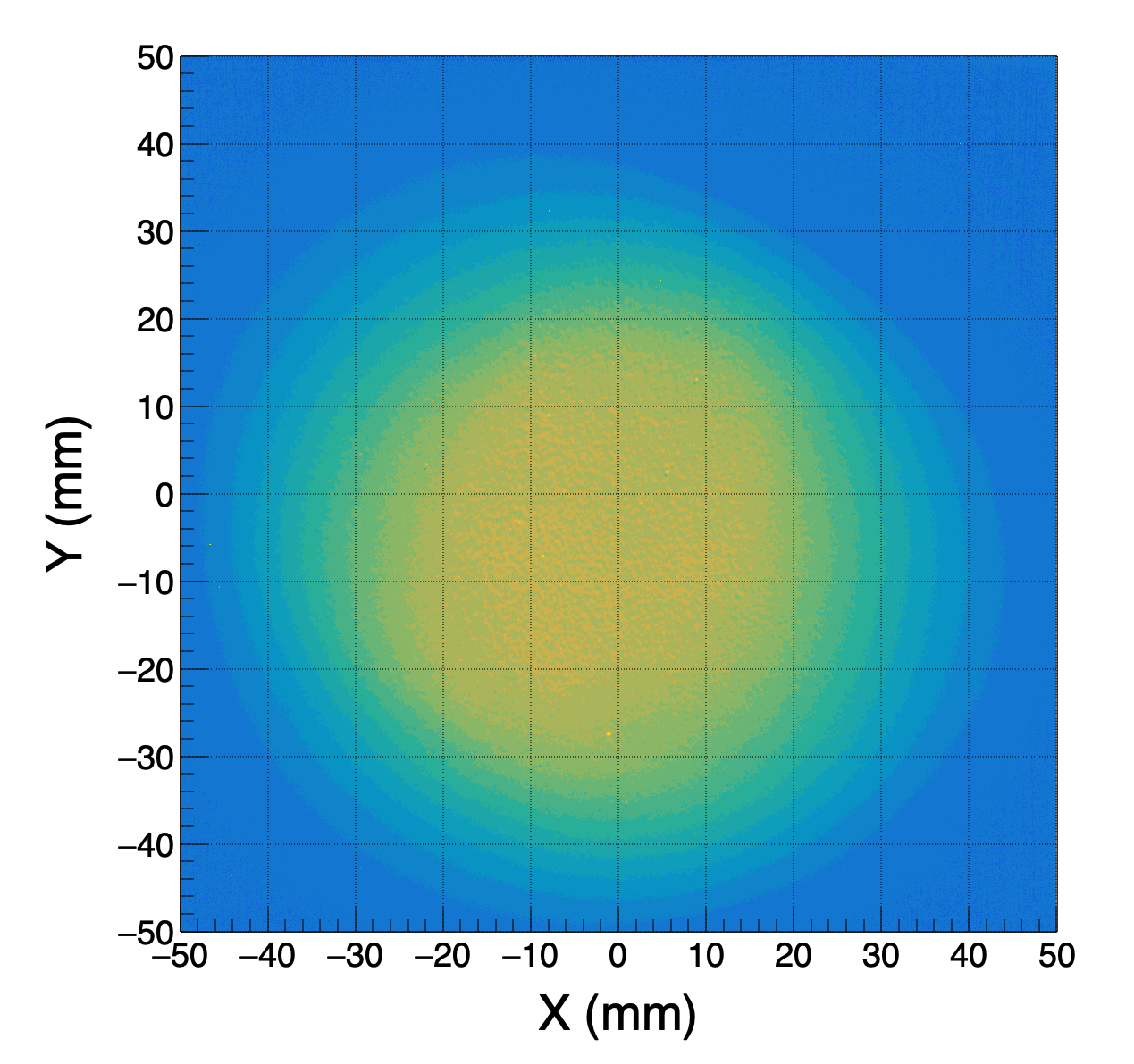}
\caption{\label{fig:flux_prof_EAR2} Beam profile in n\_TOF EAR2 obtained with the big collimator as measured with the imaging device. }
\end{figure}

Figure~\ref{fig:flux_prof_EAR2} shows the beam profile measured with the imaging device (described in Section~\ref{ssec:detection}). 

\subsection{Detection system}
\label{ssec:detection}
The imaging system used provides a single radiograph of the object under investigation, mapping the neutron beam in two dimensions perpendicular to the direction of the beam itself. To perform a complete scan of the object under investigation, the experimental station hosting it allows shifting the sample in the two directions on the horizontal plane. 

A commercially available precise optic neutron imaging system from Photonic Science has been employed for this test. It is based on an air-cooled sCMOS camera coupled with a ZnS/$^6$LiF neutron scintillator, that emits photons at 520~nm, with an active area of $100 \times 100$ mm$^2$. The scintillator has approximately 100~$\mu$m thickness, optimized for resolution. The camera sensor has an active input area of $13.3 \times 13.3$ mm$^2$ mapped by $2048 \times 2048$ pixels, resulting in an optical pixel resolution of 6.5~$\mu$m. The camera operation is entirely controlled via computer, and the adjustable features include gain, integration period, pixel clock frequency and image capture mode itself. The acquisition can be triggered by an external signal, characteristic that for a pulsed neutron beam as the one of n\_TOF allows to minimize the background coming from the experimental area and the radioactivity of the object under investigation.

\begin{figure}[ht]
\centering %
\includegraphics[width=.46\textwidth]{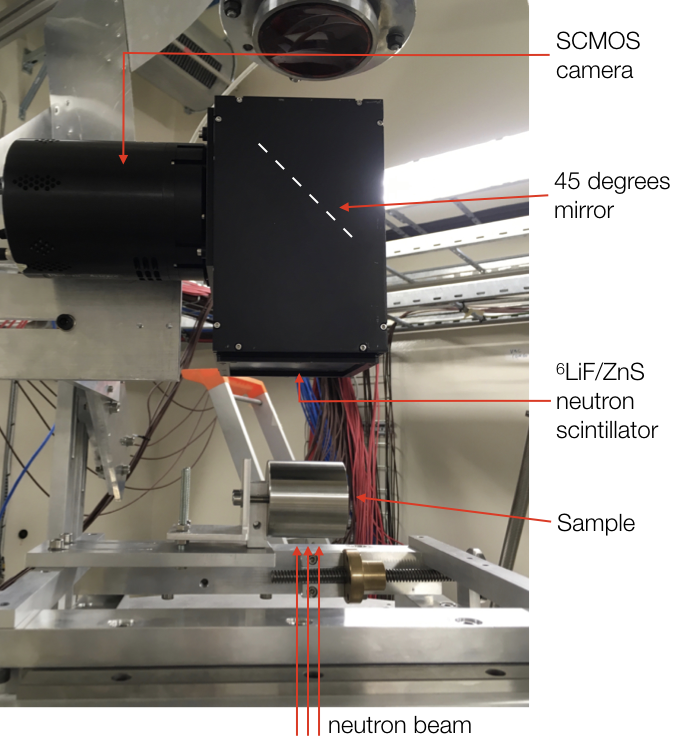}
\qquad
\includegraphics[width=.36\textwidth]{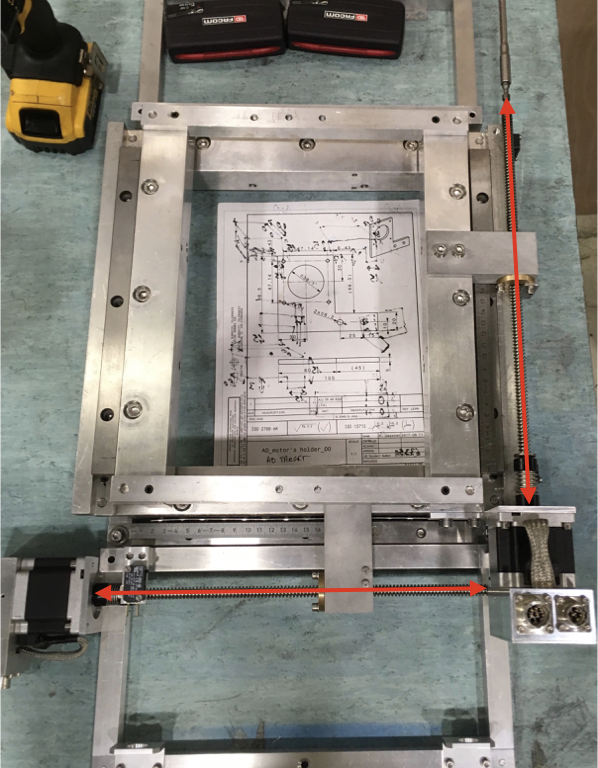}
\caption{\label{fig:imaging_device} (Left) Image of the imaging measuring setup in the Experimental Area 2, with the different parts of the setup. (Right) The sample can be moved in the plane perpendicular to the beam direction to adjust its position with a dedicated movable system.}
\end{figure}

Figure~\ref{fig:imaging_device} shows a photo of the imaging measuring setup in the EAR2, both as installed in the experimental area and a more detailed view of the sample-holding station itself.

\section{Experimental samples}
\label{sec:samples}
In order to validate the neutron imaging setup and measure its performances, five different samples have been tested within three imaging sessions in 2015, 2016 and 2017. All of these samples are related to operationally-related and R\&D activities launched at CERN for the design and manufacturing of a new set of antiproton production targets (named AD-Target)~\cite{New_ADT}. The obtained images - in addition to validating the neutron imaging technique at n\_TOF EAR2 - had a direct technical application within this project.

Antiproton production requires the use of a very high density metal as target material and a very focused and intense primary proton beam impacting onto it. This leads to a huge and fast energy deposition in the bulk of target material, with its subsequent sudden increase of temperature and appearance of violent stress waves, which can eventually fracture it~\cite{C_PhD}. This phenomena has been extensively studied both numerically~\cite{ADT_Hydro} and experimentally~\cite{HRMT27_,HRMT42}. For the latter, tests using real proton beams were carried out under the described conditions using CERN's HiRadMat facility~\cite{HiRadMatFac}. In this context, neutron imaging arises as a very convenient non-destructive technique to evaluate the damage in these high density metals after such tests, since conventional techniques as X-rays radiography cannot penetrate the target materials to reveal its internal state. In addition, the fact that the experimental area where the neutron imaging was performed (EAR2) is already categorized as a class-A lab, authorized to deal with radioactive samples and unsealed sources, simplified significantly these post-irradiation inspections, since the tested materials are considerably activated and their transport to other imaging facilities would involve formal approval of regulatory agencies. 
The geometry and characteristics of the three types of inspected samples are described in the following subsections.

\subsection{AD-Target Designs}
\label{sec:AD-T samples}

\begin{figure}[ht!]
\centering
\includegraphics[width=1\textwidth]{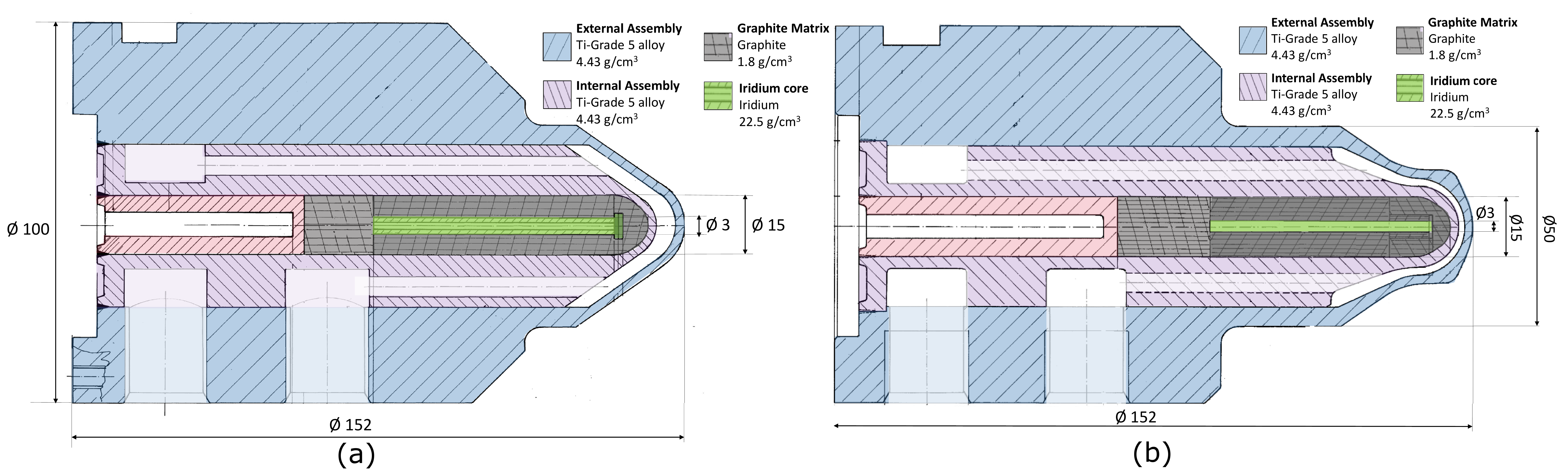}
\caption{\label{fig:ADT_draw} Simplified drawings of two different AD-Target designs inspected by neutron imaging (concept in operation since 1989 to 2018). During operation, intense and short proton pulses (coming from the left) are impacted onto their iridium core. (a) Target with non-irradiated core. (b) Radioactivated target subjected to $\mathrm{10^{6}}$ high intensity proton pulses, corresponding to roughly to $\mathrm{2\cdot 10^{18}}$ protons on target.}
\end{figure}

Figure~\ref{fig:ADT_draw} shows simplified drawings of the two inspected antiproton targets. The design and manufacturing of such targets dates back to late 1989 and correspond to the target concept employed for antiproton production at CERN since then up to 2018~\cite{ADT_Hist}. 

Even if there are small differences in the geometry of the two targets, both have in common their two-piece envelope assemblies in  titanium grade-5 alloy 100~mm in external diameter, the 15~mm diameter graphite containing matrix, and the 3 mm diameter iridium core. During operation, proton beams (coming from the left) would impact such core, leading to a sudden deposition of energy and thermo-mechanical load. 

The main difference between the inspected targets is that the one of Figure~\ref{fig:ADT_draw}-(a) was a spare never used in operation (not radioactive), while the one of Figure~\ref{fig:ADT_draw}-(b) was in operation for around 8 years, receiving up to $\mathrm{10^{6}}$ proton beam impacts.  This second target was therefore significantly radioactive (residual dose of 10~mSv/h at 10 cm) and contaminated at the surface, with its iridium core potentially damaged and/or fractured. 

The neutron imaging of the non-radioactive spare target was carried out both in the sessions of 2015 and 2016 (using a small and a big collimator respectively) while the activated one was inspected only in 2017, once the imaging methodology was well established. 

\subsection{HiRadMat-27 Experiment Targets}\label{sec:HRMT-27samples}

\begin{figure}[ht]
\centering
\includegraphics[width=0.7\textwidth]{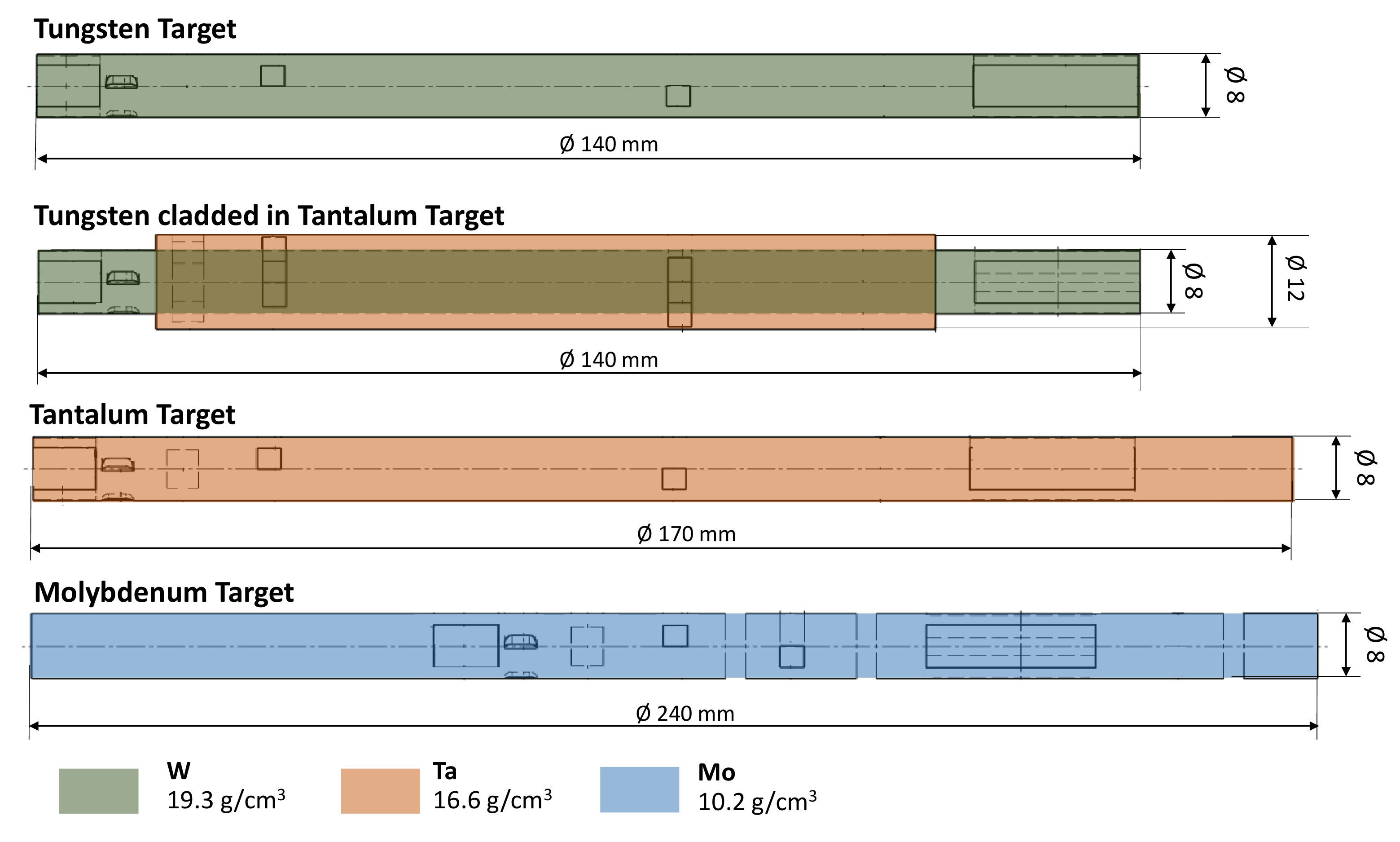}
\caption{\label{fig:HRMT27_draw} Geometry of the HiRadMat-27 targets~\cite{HRMT27_} inspected by neutron imaging in the session of 2017.}
\end{figure}

The HiRadMat-27 experiment was carried out in 2015 in order to obtain, in a controlled environment, analogous conditions to the ones taking place in the AD-Target core~\cite{HRMT27_}. Thirteen targets made of refractory metals, all potential candidates for core materials in a new AD-Target design, were exposed to proton beam impacts in order to crosscheck numerical simulations and assess their response at such conditions. Figure~\ref{fig:HRMT27_draw} shows the geometry of some of these targets which, after the experiment and before other destructive inspections, were brought to EAR2 for neutron imaging. In this way, internal cracks induced by the proton beam could be identified, thus aiding in the selection of relevant areas to cut for further studies.

As shown in the figure, most of the inspected targets were 8 mm in diameter with densities from 10.2 to 19.3 $\mathrm{g/cm^3}$. In addition, one of the targets was an 8 mm diameter tungsten rod cladded in 2 mm thick tantalum, leading to an external diameter of 12 mm. During the experiment, these targets were subjected to a relatively low number of proton beam pulses, therefore their residual doses were only in the order 30 $\mu$Sv/h and no remote manipulation was necessary. 

\subsection{HiRadMat-42 Experiment Target}\label{sec:HRMT42sec}

\begin{figure}[ht]
\centering
\includegraphics[width=0.8\textwidth]{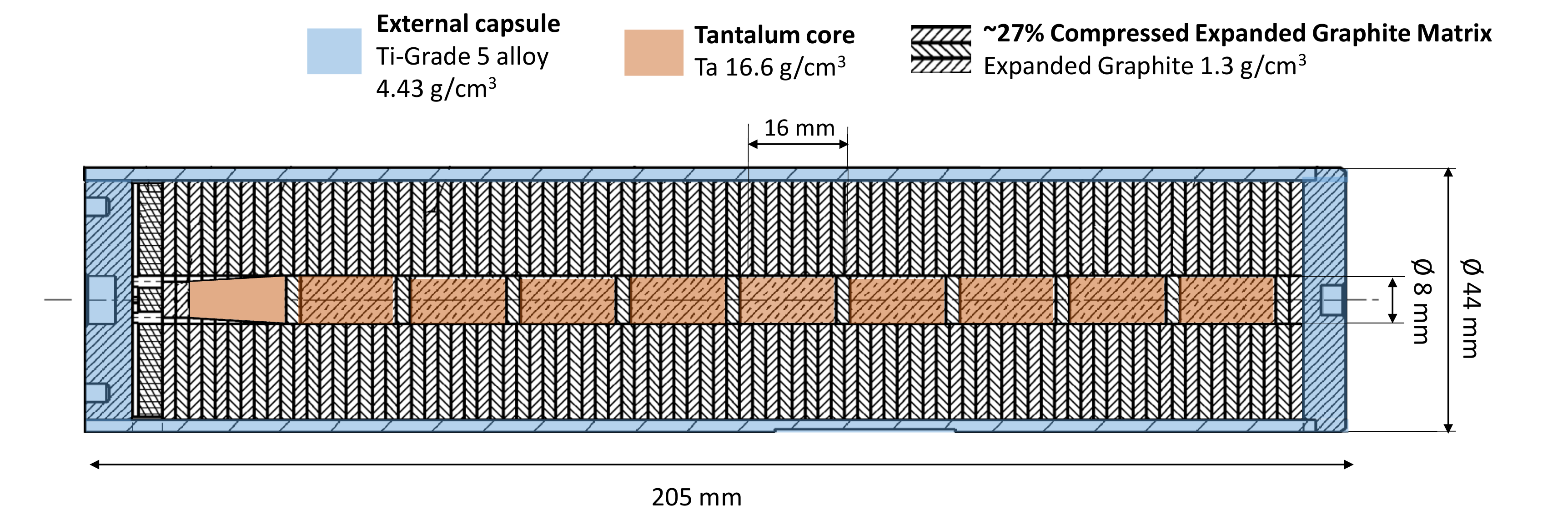}
\caption{\label{fig:HRMT42_draw} Geometry of the HiRadMat-42 target inspected by neutron imaging in the session of 2017~\cite{HRMT42}.}
\end{figure}

Similarly to the HiRadMat-27 experiment, the HiRadMat-42~\cite{HRMT42} experiment was performed in 2017 for testing a first, simplified and scaled, prototype of the new AD-Target design, shown in Figure~\ref{fig:HRMT42_draw}. This target consisted of 10 rods of tantalum 8 mm in diameter and 16 mm in length embedded in a matrix of compressed expanded graphite and encapsulated in a titanium grade-5 alloy. Differently from the HiRadMat-27 experiment (in which only a few pulses per target were impacted), this target was exposed to 50 proton beam impacts, aiming at investigating progressive material damage in the Ta core. During the neutron imaging at n\_TOF, the dose rate at contact of this target was in the order of 300 $\mu$Sv/h. A full neutron tomography of this target was also performed at the PSI-SINQ NEUTRA beam line~\cite{Neutra}, with results presented in Ref.~\cite{HRMT42}.

\section{Experimental results of neutron imaging at EAR2}

As already introduced, the neutron imaging method has been validated in three separate campaigns in 2015, 2016 and 2017, using the small (70 mm) collimator in 2015 and the big (96.8 mm) in 2016 and 2017 respectively. 

\begin{figure}[htb]
\centering 
\includegraphics[width=.42\textwidth]{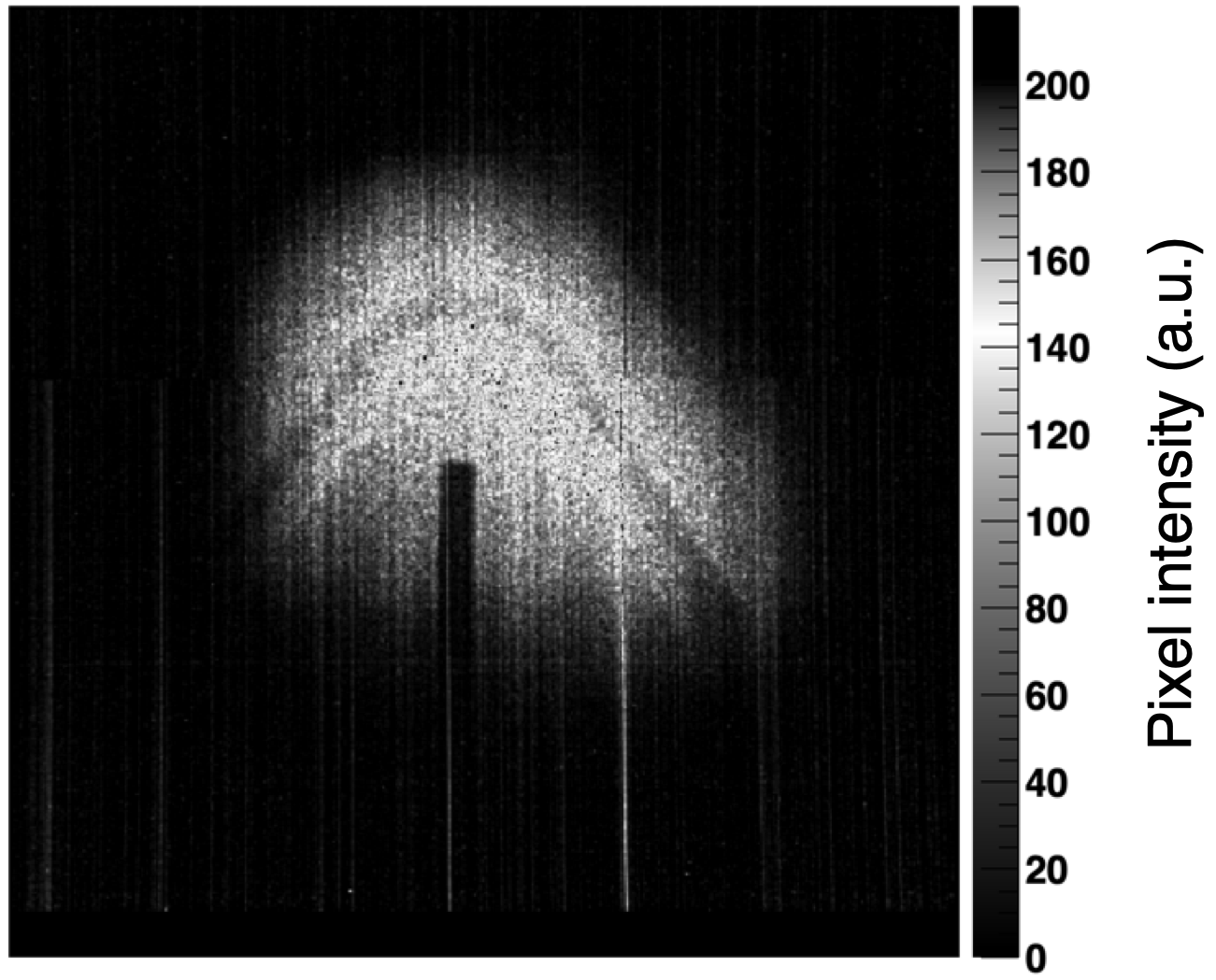}
\qquad
\includegraphics[width=.44\textwidth]{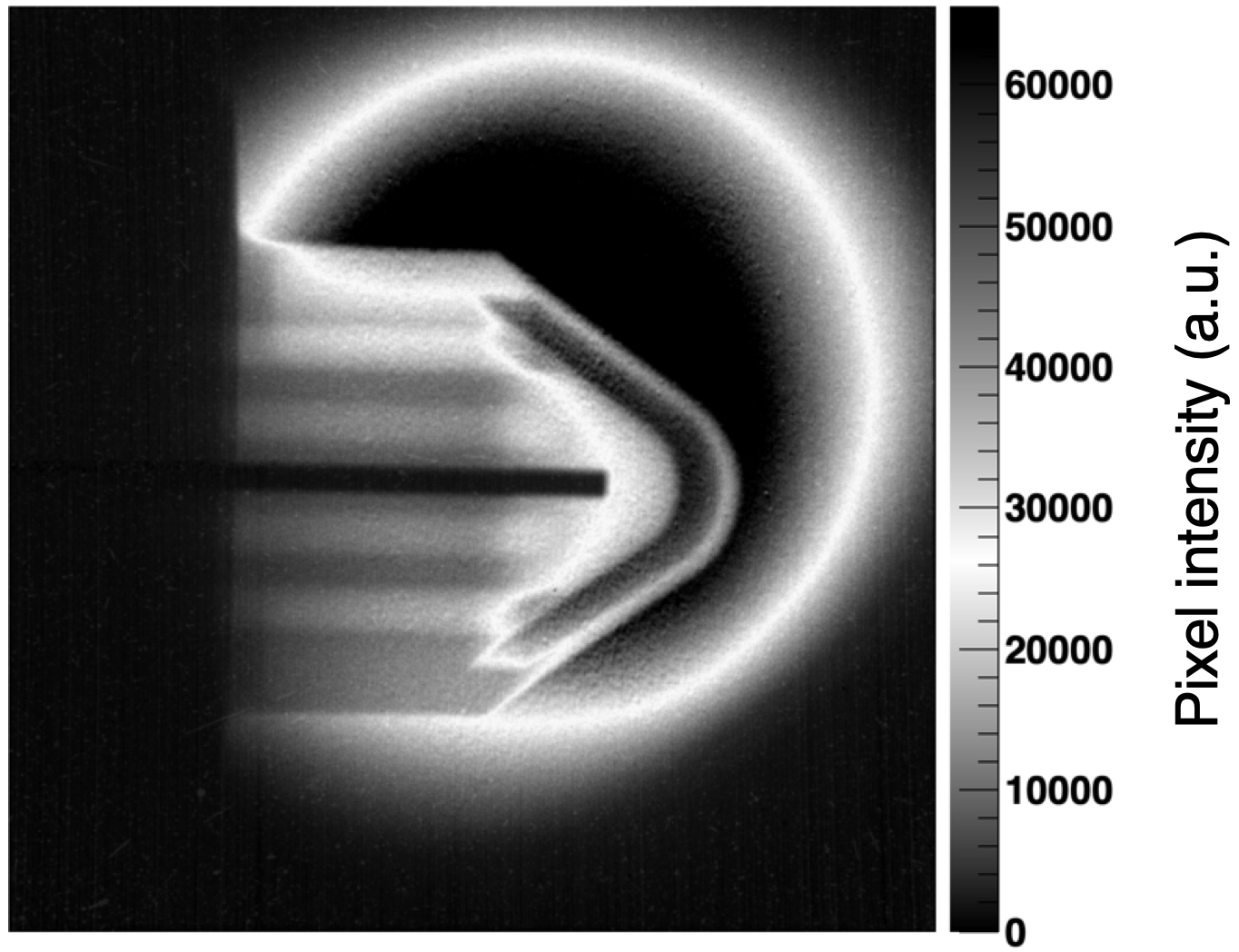}
\caption{\label{fig:tests} Comparison of the image of a spare non-irradiated AD-Target obtained at the n TOF EAR2 imaging station for the small (left panel) and big (right panel) collimator.}
\end{figure}

During the tests in 2015 and 2016, the non-irradiated AD-Target introduced in Figure~\ref{fig:ADT_draw}-(a) has been investigated. Results are shown in the two panels of Figure~\ref{fig:tests} (left for the small and right for the big collimator, respectively). As it can be seen, in both cases the inner iridium core of the target is perfectly visible with a very good contrast, which proves the feasibility of the technique. While with the small collimator the spatial resolution is better due to the lower $D/L$ ratio, the advantage in using the big collimator is a much larger homogeneously irradiated area, with a higher neutron intensity and improved contrast.

Following the successful proof of concept, the irradiated samples described in Section~\ref{sec:samples} have been inspected during the campaign of 2017 with the goal to examine their inner structure and to identify potential material damage. In this respect, the geometrical and material constraints of the samples played an important role in planning the beam-time per radiograph. In fact, for the HiRadMat-27 targets that did not present any encapsulation and for the HiRadMat-42 target, with a more complex design but still light enough, statistics of about $2\times10^9$ thermal neutrons/cm$^2$ - resulting in an average of 45 minutes per radiograph - was sufficient to obtain the desired contrast. On the contrary, the old AD-Targets shown in Figure~\ref{fig:ADT_draw} consist of a denser assembly where the inner iridium core is embedded in the graphite matrix and the titanium alloy envelopes (up to 100 mm of external diameter). In this case an average of $6.5\times10^9$ thermal neutrons/cm$^2$ - for a measuring time of 2 hours and 40 minutes - were accumulated per radiograph to achieve the desired contrast.

\begin{figure}[ht]
\centering
\includegraphics[width=.53\textwidth]{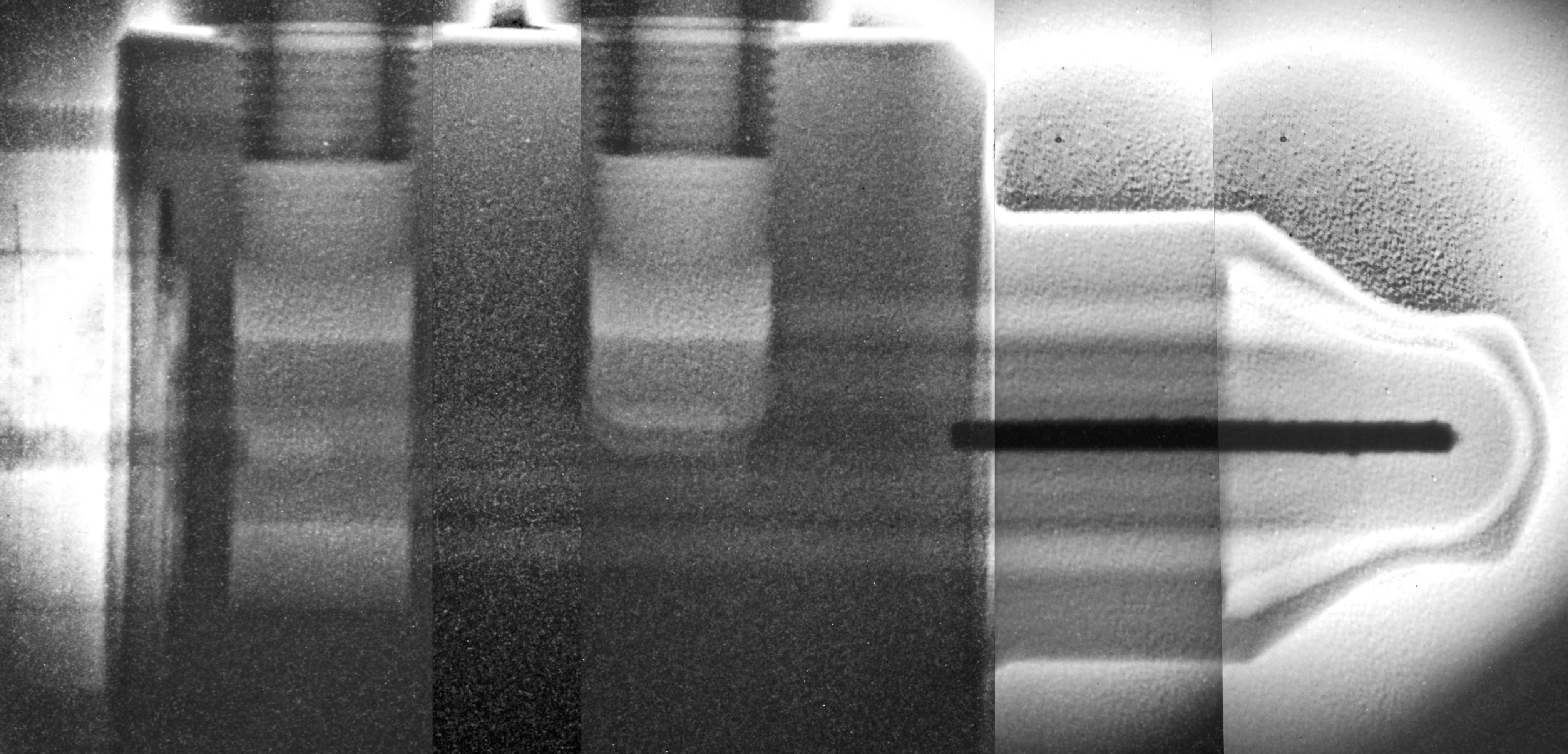}
\quad
\includegraphics[angle=90,origin=c,width=.33\textwidth]{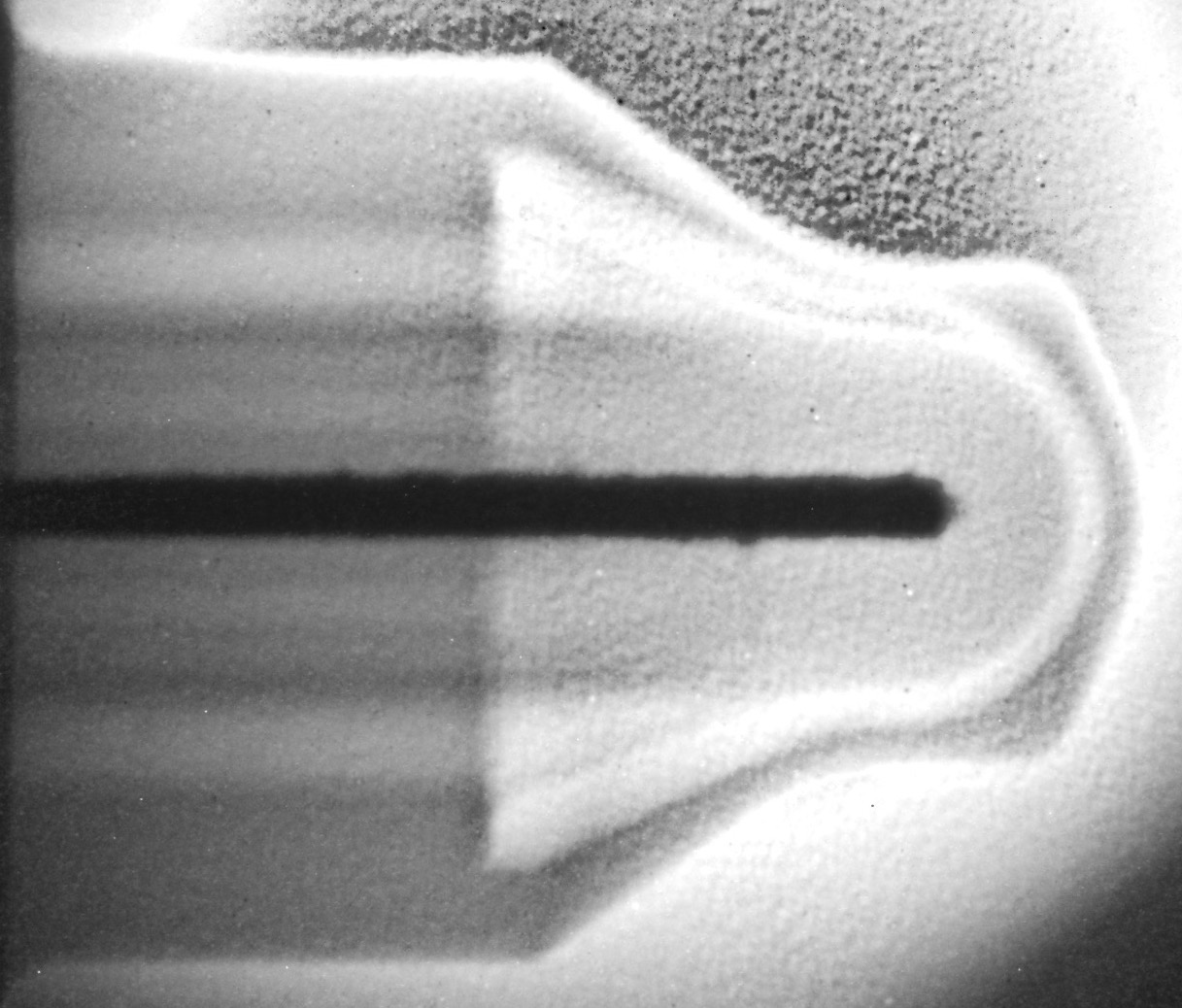}
\caption{\label{fig:AD} Radiograph of an irradiated AD-Target assembly. From the zoom in the right panel, the non-uniformity of the inner core (which should present itself as a perfect straight line, as in Figure~\ref{fig:tests}) is clearly visible highlighting internal damage.}
\end{figure}

\begin{figure}[ht]
\centering
\includegraphics[width=.53\textwidth]{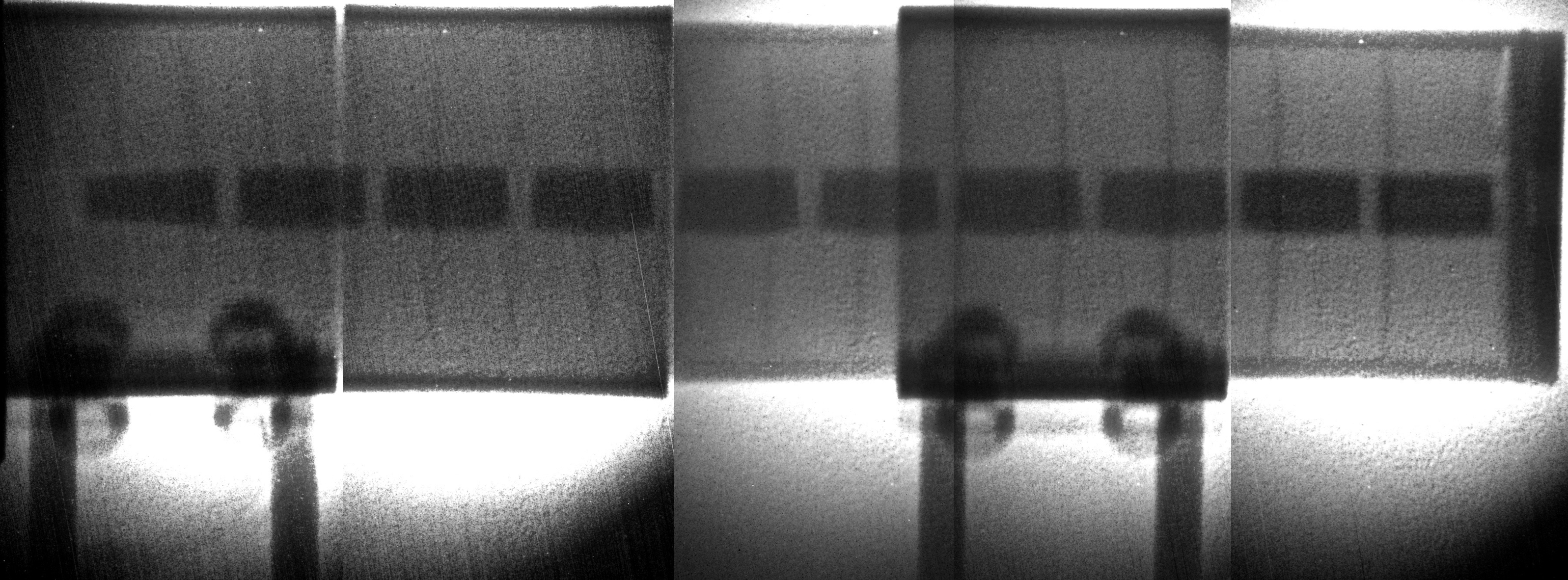}
\quad
\includegraphics[angle=90,origin=c,width=.33\textwidth]{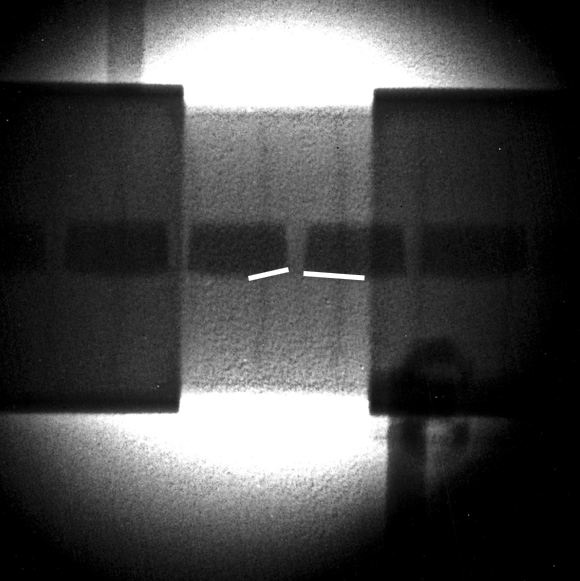}
\caption{\label{fig:HRMT42} Radiograph of the HiRadMat-42 target assembly. The sliced tantalum cores appear bent due to permanent plastic deformation induced by the proton beam impacts, as also confirmed by the neutron tomography at the NEUTRA neutron beam line (PSI)~\cite{HRMT42}.}
\end{figure}

\begin{figure}[ht]
\begin{center} 
\includegraphics[width=1.\textwidth]{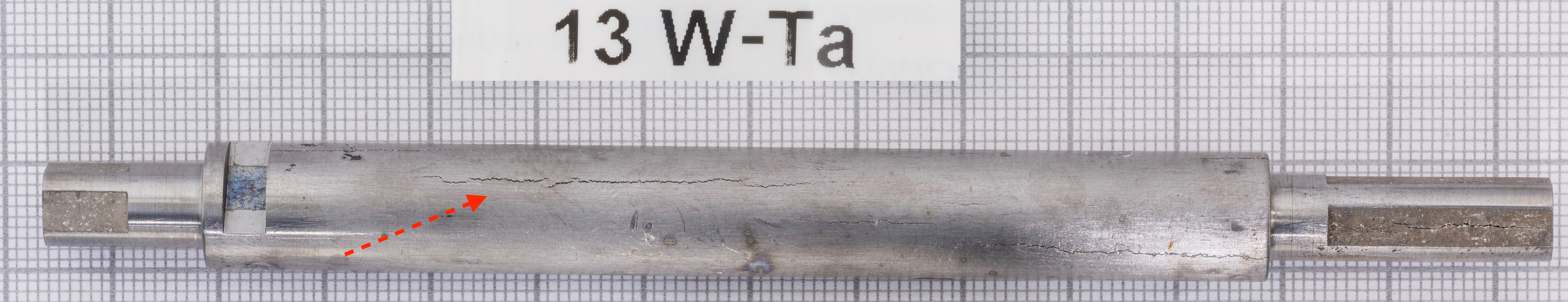}
\includegraphics[width=.8\textwidth]{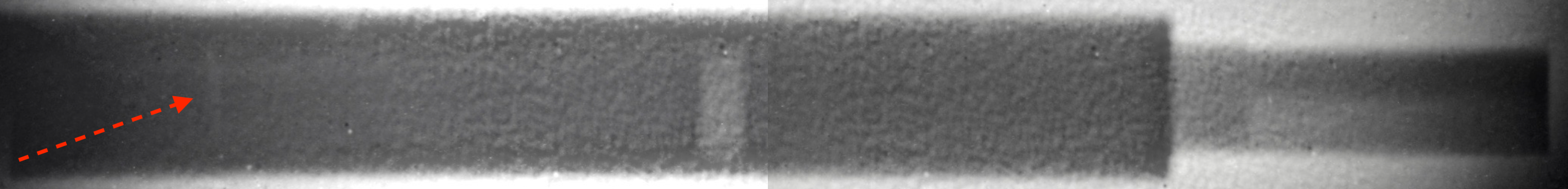}
\caption{\label{fig:HRMT27} Radiograph of the HiRadMat-27 Ta-cladded W rod (bottom panel). The red arrows highlight two big cracks forming a cross-like structure, of which the vertical one is not visible from the outside. In this case, the radiograph highlighted in fact a crack in the inner W-core.}
\end{center}
\end{figure}

\begin{figure}[htb]
\centering
\includegraphics[width=.45\textwidth]{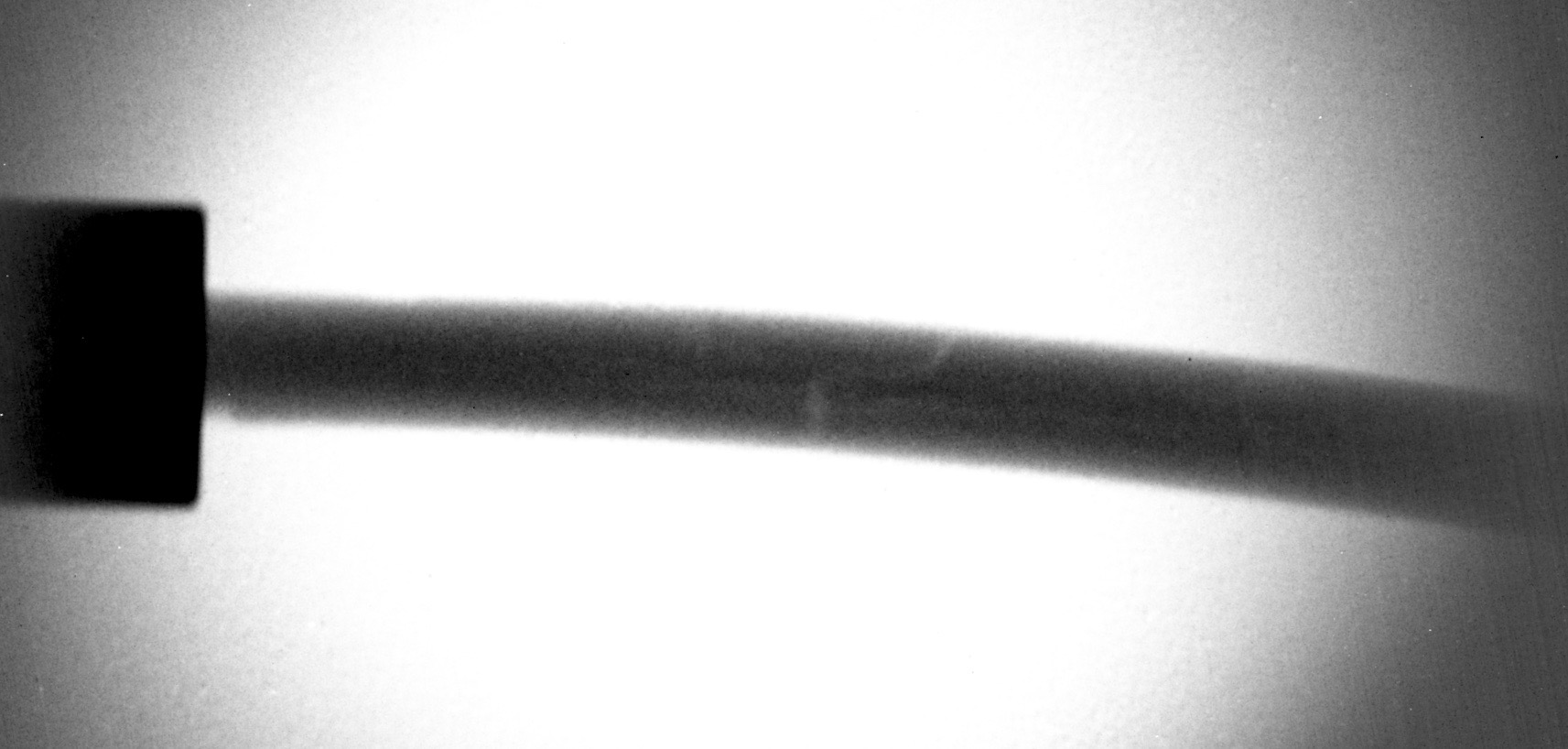}
\qquad
\includegraphics[width=.45\textwidth]{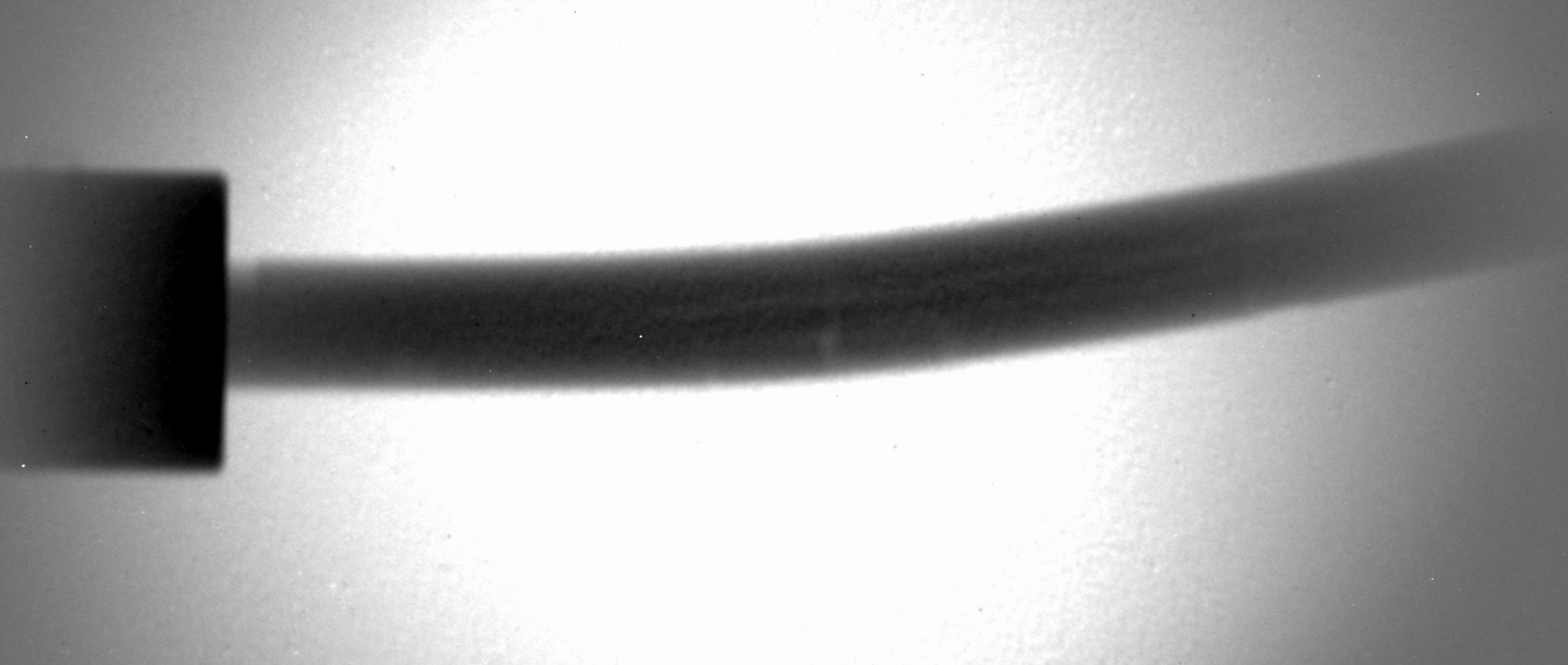}
\caption{\label{fig:HRMT27_rot} Radiographs of one of the HiRadMat-27 W rods before (left panel) and after (right panel) a rotation of 120$^\circ$. Different fractures are visible depending on the orientation of the rod with respect to the neutron beam.}
\end{figure}

The results obtained are presented in Figures~\ref{fig:AD} to \ref{fig:HRMT27_rot}. Since, as described in Section~\ref{ssec:beam_par}, the beam spot was 9 cm in diameter, to obtain a full radiograph of the objects under investigation several pictures had to be taken with the support of the movable table, namely six for the AD-Target, seven for the HiRadMat-42 target, and two for each of the HiRadMat-27 targets. 

As can be seen from the figures, the higher neutron fluence obtained with the big collimator allows one to clearly distinguish and investigate the inner structures of the samples presented in Section~\ref{sec:samples}. For instance, Figure~\ref{fig:AD} shows the neutron-radiograph of the previously proton irradiated AD-Target. In particular, the right panel of the figure distinctly shows a non-uniformity in the contour of the iridium core, which was not present in the ones obtained from the non-irradiated spare AD-Target (Figure~\ref{fig:tests}). This non-uniform contour along the core of the proton irradiated AD-Target is therefore a clear indication of its fragmentation and internal damage due to the extreme stress waves induced by the proton beam impact during operation, as suggested by numerical simulations in Ref.~\cite{ADT_Hydro,C_PhD} and verified after target opening (detailed in a forthcoming paper).

In Figure~\ref{fig:HRMT42}, the inner tantalum rods of the HiRadMat-42 target assembly experienced permanent plastic deformation and are slightly bent due to excitation of bending modes induced by the thermo-mechanical load of the proton beam impact. This conclusion has been confirmed by the additional tomography performed at the NEUTRA line in the SINQ facility in PSI~\cite{HRMT42}, as well as after the subsequent target opening. The results of the neutron tomography performed at PSI indicated that the Ta cores suffered internal voids (up to 0.8~mm size)~\cite{HRMT42}. Unfortunately, the voids cannot be observed in the single neutron imaging of figure~\ref{fig:HRMT42} since a higher spatial resolution would be necessary to be sensitive to the minimal absence of Ta material in these voids.

Regarding the HiRadMat-27 targets, a more significant result has been obtained for the tungsten target cladded in tantalum. As shown in Figure~\ref{fig:HRMT27}, two major fractures have been detected, one longitudinal and one transverse. The longitudinal one reached the surface and is also externally visible, while the transverse one is enclosed in the tungsten core and is only revealed by the radiograph. 

Another interesting result obtained with the HiRadMat-27 targets is the expected angle-dependence of the method, meaning that the radiographs showed different internal structures depending on the orientation of the sample with respect to the neutron beam. This behaviour can be observed in Figure~\ref{fig:HRMT27_rot}, where two radiographs of one of the tungsten targets are shown before and after a rotation by 120 degrees. The fact that different cracks are detected in the radiographs depending on the orientation is clearly highlighted. This paves the way to potential applications of neutron tomography techniques at n\_TOF EAR2 imaging station.

\section{Conclusions and perspectives}
The unique characteristics of the n\_TOF facility at CERN have been successfully exploited to perform neutron radiography on different irradiated materials and target assemblies. Following the results from three feasibility tests, the experimental setup has been chosen in order to maximize the contrast (i.e. the neutron fluence) and the beam spot, using the bigger collimation system available in the facility. In this configuration, in 2017 several targets -- previously irradiated in the HiRadMat CERN facility -- have been investigated, analyzing their inner structures and qualitatively assessing their damage.

The successful and promising results obtained in the 2017 measurement campaign could now be used as a starting point to further develop the imaging capabilities of the n\_TOF facility. In particular, a new, dedicated, collimation system is being designed to improve the achievable resolution preserving at the same time the high neutron flux necessary for a good imaging contrast. In addition, the measuring station will be equipped with a remote-controlled rotating system in the z-axis, to make it easier to take radiographs at different angles and to improve the flexibility of such studies (e.g. neutron tomography).

Besides the facility-related improvements, the detection system is also foreseen to be upgraded with a more solid scintillator-optics assembly. In particular, the ZnS was found degrading its time characteristics with the exposure to the high neutron flux -- also at neutron energies above hundreds of keV -- of the n\_TOF facility.

New detection setups are under investigation in order to exploit the time-of-flight technique, characteristics of the facility. In particular, the possibility of coupling the standard imaging system with a fast scintillator to use the Resonance Shape Analysis from the (n,$\gamma$) reaction to identify the material composition of the object under investigation. In addition, the use of fast neutron detectors, as for example Micro Channel Plates, could open the possibility of taking material-selective radiographs if the acquisition window of the camera is gated on a resonance specific of a given isotope. 


\vspace{6pt} 




\funding{This research received no external funding.}

\acknowledgments{The authors would like to thank the support of the n\_TOF Collaboration for the execution of this work as well as the Engineering Department, Sources Targets and Interactions Group responsible for CERN's Beam Intercepting Devices. 
One of the author (MC) would like to thank A. Pietropaolo and F. Grazzi for the initial discussions aiming at proposing and setting up the neutron imaging station at n\_TOF. The authors kindly acknowledge the support and technical exchanges with the PSI NEUTRA team in establishing the imaging station as well as in the execution of the tomographies at PSI, namely M. Strobl, E. Lehmann and C. Gr\"unzweig.}

\conflictsofinterest{The authors declare no conflict of interest.} 

\externalbibliography{yes}
\bibliography{ref}



\end{document}